\documentstyle[11pt,newpasp]{article}

\begin{document}

\title{Theoretical perspectives }

\author{E. Athanassoula }
\affil{Observatoire de Marseille, 2 place Le Verrier, 13248 Marseille
cedex 04, France}




\begin{abstract}
I first comment on some recent advances in
computing equipment for CPU-intensive numerical simulations,
and on possible developments in the near future. I then discuss some
particularly important and yet
unsolved problems in dynamics and
evolution of galaxies, on which analytical, numerical and 
observational effort should be focused.
\end{abstract}




\section{Introduction}

In the case of observations it is easy and meaningful to talk about
perspectives. Plans are made for new satellites, new telescopes and new 
instruments, and from their specifications one can make an educated guess
about what new observations will be made and extrapolate to 
what new information they will bring us (knowing well that some
surprises might be in store as a bonus). This is not the case for
theory, the progress of which mainly depends on bright new ideas, 
which are of course impossible to predict.

Theoreticians, however, have one tool, the computer, whose progress
over the past few decades has been tremendous, and about whose future
advances it 
is possible to make some predictions. This is true even for personal
computers or workstations, but particularly so for machines on which
one can perform CPU-intensive numerical simulations. I will thus
devote the next section to these types of machines. I
will then briefly discuss some theoretical problems of particular interest, 
on which important progress could be made in the next few years,
particularly with the help of numerical simulations.

\section{ Computers for CPU-intensive calculations }

The evolution of computers over the last half-century has been
amazing, and the numerical simulations it allowed have been the source of
important progress in galactic dynamics. Very large, CPU-intensive
calculations are possible on mainly three types of computers, whose
advantages and disadvantages will be considered in turn.

\subsection{Supercomputers } \label{supercomputers}

Supercomputers are facilities which are either national or at least
institutional. As such, they are run by an operating team and the user 
does not have to worry about hardware maintenance. They also provide
good computer libraries and manuals, greatly facilitating the
programming task, while the operating team is often available for
advice. Furthermore they often have very large memories. They can, in
principle, be used for a very large variety of 
programs. Finally the rapid recent increase in communication
speeds has greatly facilitated the use of these facilities when 
not in-house.

As disadvantages let us mention the large purchase and running 
cost, the relatively small flexibility of use (one has to make
proposals at given deadlines, sometimes far in advance) and the fact
that the software that 
is tailored for them, since it is largely based on their specific
libraries, is non-portable.
As a consequence in many countries they start to be phased out in
favour of smaller, more dedicated machines, and
this tendency will probably be accelerated in the future.

\subsection{Beowulf-type systems } \label{beowulf}

Beowulf is a name commonly given to a computer consisting of a large
number of PCs, coupled by a dedicated and fast network (cf. www.beowulf.org).

Their relatively low price makes it possible for small institutions or 
departments to acquire them, provided that some engineering personnel
is available, or that a few astronomers are ready to invest some of
their time. They are somewhat more difficult to program on than
supercomputers, since they do not have as efficient libraries,
but this is often compensated by their in-house availability and their
very good price-to-performance ratio. Furthermore software written for
one such system can be relatively easily used on any other.

It is thus easy to predict that such systems will become more and more 
frequent, and reach ever-increasing performances due to the amazing
advances in PC technology.

\subsection{GRAPE systems } \label{grapes}

GRAPEs - for GRAvity piPEs - are special purpose boards on which is
cabled the most CPU-consuming part of N-body calculations, namely the
calculation of the gravitational force. They
are coupled to a standard workstation via an Sbus/VMEbus, or a PCI bus 
interface. The host computer provides the GRAPE with the masses and the
positions of all the particles, and the GRAPE calculates and returns the
accelerations and the potentials. These boards are developed by a group in
Tokyo University, headed initially by D. Sugimoto, and now by
J. Makino. The history of the GRAPE project, starting more than 10
years ago with GRAPE-1, is given by Makino \& Taiji (1998). There are
essentially two families of GRAPEs, those with odd numbers, that have
limited precision, and those with even numbers, which have high
precision.

\subsubsection{GRAPE-5 } \label{g5}

The latest arrival in the family of the odd-numbered GRAPEs is GRAPE-5 (Kawai 
et al. 1999), and it follows to a large extent the architecture of
GRAPE-3. As all other GRAPEs, it calculates the forces and 
potentials from a set of particles, and also gives the list of 
nearest neighbours, particularly useful when doing 
SPH or sticky particle calculations. It has a peak performance of
38.4 Gflops per board and a clock frequency of 80 MHz. Each board
has 8 processor chips, and each chip has 2 pipelines. It is coupled
to its host computer via a PCI interface.

GRAPE-5 is a vast improvement with respect to GRAPE-3. It is 8 times
faster and roughly 10 times more accurate. The communication speed has 
also improved by an order of magnitude, while the size of the neighbour list is
considerably lengthened, so that it can hold up to 32768 neighbours for 
48 particles, thus rendering particle-hydro
simulations much easier to program. At the time this talk was given,
only the prototype GRAPE-5 had been tried out. As I am writing these
lines several GRAPE-5 boards are already in use
both in Komaba (Tokyo University) and the Observatoire de Marseille,
while several more groups make plans to acquire them.
Tokyo University has plans for building a massively parallel GRAPE-5
system with a peak performance of about 1 Tflops. On such a system one 
treecode time-step for 10$^7$ particles should take about 10 secs.

\subsubsection{GRAPE-6 }

GRAPE-6 will be the successor of GRAPE-4, whose architecture it is
basically following. It calculates not only the potential and the force,
but also the derivative of the force, thus allowing the implementation 
of individual time-step schemes (e.g. Makino \& Aarseth 1992). A
single GRAPE-6 chip should be the CPU equivalent of a whole GRAPE-4
board. The chip is presently in
its testing phases, and should be commercially available by 2001. Both 
single chip units (baby-6), and single board units (junior-6, with 16
chips) are planned.

\subsubsection{PROGRAPE-1 }

In particle
hydrodynamics, GRAPEs are used only to calculate the
gravitational forces and the list of nearest neighbours. The
actual evaluation of the SPH interactions is done on the host
computer, thus hampering the performance. 
Nevertheless building a special purpose SPH machine
may not be a good idea, since there are a large number of varieties of 
particle hydrodynamics,
and each would necessitate its own GRAPE implementation.
It is thus preferable to have recourse to reconfigurable
computing, or field-programmable gate arrays (FPGA). Such chips,
also called programmable chips, 
consist of logic elements and a switching matrix to connect them, and
their logic can thus be reconfigured.

In order to reduce both the work of the designer and that of the
application programmer, PROGRAPE is 
specialised to a limited range of problems, namely the evaluation of
particle-particle interactions. The application programmer 
has to change only the functional form of the interaction. It is thus in 
a way intermediate between the standard GRAPE systems and general
purpose computers. Another project for SPH FPGAs is
being developed in a collaboration between groups in Heidelberg and Mannheim.
The Tokyo group, after completing PROGRAPE-1 (Hamada et al. 1999), is
now starting on 
PROGRAPE-2, a massively parallel extension of PROGRAPE-1, which should 
achieve somewhere between 1 and 10 Tflops, and be available in a
couple of years.

\subsubsection{Advantages and disadvantages of GRAPE systems }

GRAPE systems are of course limited to N-body type simulations, and
thus should not be purchased by groups having other types of CPU-intensive
calculations. One of their big advantages is that they are within the
reach of a small group or department, while their availability makes 
it possible to envisage long-term projects. To this one should
add their excellent price-to-performance ratio, as witnessed by the two 
Gordon Bell prices they have won so far. Finally users of GRAPE
facilities form a small community with close links, discussing their
hardware and software environments, helping each other along, and
often exchanging software. For all these reasons I wholeheartedly
recommend GRAPE systems to groups which perform CPU-intensive N-body
simulations and have a sufficient level of computer knowledge. 

We have thus seen that both beowulf-type systems and GRAPEs have
important advantages. The
choice between the two depends basically on the type of applications
(mainly N-body or a broader 
spectrum) and on personal preference. It is not, however, necessary to
chose between the two, since it is possible to envisage a beowulf-type
system with GRAPE boards attached to some or all of its nodes. On a
similar line the National Observatory of Japan in
Mitaka has plans for connecting sixteen GRAPE-5 boards to a supercomputer.

\section{Problems of particular interest}

\subsection{Dark matter}

Although dark halos have been with us for over twenty years, there is
still a lot we do not know, or do not understand about them. They were 
first introduced in the seventies by Ostriker \& Peebles (1973) as a
way of stabilising discs against the ubiquitous bar
instability. Today it is understood that they can achieve this, in
the linear regime, only if they are sufficiently concentrated to cut
the swing amplifier cycle (Toomre 
1981), or, in the non-linear regime, to prohibit the incoming
waves from tunneling through to the center of the galaxy. Although
the extent and amount of mass in the outer halo is irrelevant to the
bar instability, it is crucial for a lot of other dynamical
issues.

Even in disc galaxies, where HI extended rotation curves have shown
clearly the necessity of an extended dark matter halo {\footnote
{Unless one allows for a modified gravity}} (e.g. Bosma 1981), there
are still a number of  
unanswered questions. One of the most crucial ones concerns the disc-to-halo
mass ratio in the main body of the galaxy. Are discs maximum? Or are they of 
relatively low mass, their dynamics to a large extent dominated by the 
massive halo? Several arguments, both theoretical and observational,
have been advanced, and yet the answer is still not clear. Thus
for example, if discs were not sufficiently heavy 2-armed
structures could not form in them (Athanassoula, Bosma \& Papaioannou
1987), while bars would decelerate relatively fast and end up as slow
rotators (Debattista \& Sellwood 1998), in both cases contrary to 
observations. On the other hand measurements of velocity dispersions in 
discs (e.g. Bottema 1993; see also Bosma in these proceedings),
favour non-maximum discs, arguing that  
massive discs would lead to very low values of the Toomre Q
parameter (Toomre 1964). Further arguments based on the 
Tully-Fischer relation come against the maximum disc hypothesis
(e.g. Courteau \& Rix 1999). Finally cosmological N-body simulations
predict less than maximum discs (e.g. Navarro 1998). How can all these
be reconciled? Are galactic 
discs maximum or not? Certainly more work is necessary here to better
understand the effect of halos on the 
dynamics of disc galaxies, and thus their masses.

\subsection{Evolution of galaxies }

Recent observations with the HST, and in the future with the NGST,
and with large ground-based telescopes, provide us with information on
the properties of galaxies at high $z$. We now 
know more about both their morphology and their kinematics.
As implied by the title of this conference, it is one of our main tasks
to understand how the morphology and dynamics of galaxies changes with 
time. As long as such observational data did not exist, the only
constraint on evolutionary scenarios was that they had to match 
observations at $z=0$. Observations at  higher redshifts
make the work of theoreticians more daunting and at the same time more
interesting.

For example Abraham et al. (1999)
argued that very few barred galaxies can be found at high
$z$. Since interactions drive bar formation (Noguchi 1987, Gerin,
Combes \& Athanassoula 1990), wouldn't it be reasonable to expect more 
bars at higher redshifts? Several answers can be proposed. One
possibility would be that at higher redhifts discs had lower
surface densities (since their mass can be assumed to grow in time
until its present level). In that case multi-armed structures
would be favoured over 2-armed ones. Since such patterns have
necessarily inner Lindblad resonances and a small extent between their 
inner and outer such resonances, one would expect fragmentary
multi-armed episodes, driven by interactions, rather than bars, in
good agreement to  
observations at higher $z$. This suggestion merits further work, 
which, together with other scenarios, would lead to a better
understanding of the morphology of disc galaxies at high redshifts. 

\subsection{Dynamics of bars }

The life of a
bar has several episodes: its formation, evolution, possible destruction and 
perhaps regeneration. All have parts which are poorly
understood, but this is particularly true for the third and, even more, 
the fourth episode.

A bar may be destroyed by the infall of a companion on its
host disc (Pfenniger 1991, Athanassoula 1996b). Furthermore
bars in discs with a gaseous component are known to commit suicide by
pushing gas towards their center, where a central concentration can form, 
destroy the orbits that support the bar and hence the bar
itself. N-body simulations (e.g. Friedli \& Benz 1993)
show that this occurs on a time-scale of the order of a few bar
rotations, i.e. that bars in discs containing gas 
should be relatively short lived. On the other hand observations show
that strong 
bars are present in over a third of all discs, and weaker ones in yet
another third, if not all the remaining discs. How can these two be
reconciled? It is of course possible, although highly unlikely, that
all bars have formed only a few rotations ago. It is also possible that we are
witnessing a second generation of bars, although this solution may have its own
problems, as will be shortly discussed below. Finally it is 
possible that SPH simulations, which have clearly illustrated this
third phase in the lifetime of a bar, give shorter
time-scales for the gas inflow, and hence for the bar destruction,
then what is the case in real bars.  

The fourth episode in the life of a bar, namely its possible
regeneration, is even less well 
understood. The disc of the galaxy, as left after the bar destruction,
is a hostile environment for a new bar to form. It is hot,
since its stars have been heated by the previous bar, and it may have
a large central concentration or bulge. How can a bar form 
in such circumstances? Two suggestions have been made so far. Sellwood 
\& Moore (1999) suggested that freshly infalling gas may cool the disc
sufficiently to allow the generation of a new bar, while 
Miwa \& Noguchi (1998) use a very strong external
forcing. Are the properties of these second generation bars, different 
in any way from those of the first generation bars? The simulation 
of Miwa and Noguchi argues that bars driven by a very strong external forcing
should rotate slower
than the spontaneous ones and end near their inner Lindblad
resonance. Seen the contradiction with observations of early type
galaxies, some further such simulations should be 
carried out, partly to see how general this result is, how much it
constrains second generation bars, 
but also in order to understand the orbital structure 
in such bars.

Bars are particularly interesting from a dynamical point of
view. There is thus a large number of further questions to be
examined. What is the fraction of chaotic orbits in self-consistent 
bars, and, more generally, the relative importance of the different types
of orbit families? What are the differences between the properties of bars
in early and late type galaxies and what are they due to? How do bars
within bars form and 
evolve? These are only few of the most interesting questions in this context. 

\subsection{Galaxy interactions and mergings. Dynamical effects on
galaxies in groups and clusters } 

Although a considerable effort has been put lately in this very
interesting topic (e.g. Barnes 1998 and references therein),
still a lot remains to be done. For example we need to understand
better interactions and mergings which are more characteristic of
higher redshifts,
e.g. by using smaller and more gas-rich discs. We also need to know more on
mergings of unequal sized galaxies (for some preliminary results see
e.g. Barnes 1998 and Athanassoula 1996a,b), an area hitherto 
insufficiently explored, since a fully 
self-consistent treatment of such cases requires considerably more
particles than equal mass interactions and mergers. Finally most simulations
have so far considered the interaction and merging of two unbarred
discs. Now that this is getting  
somewhat better understood we should consider cases in which at least one of
the partners is barred (Athanassoula, 1996a, 1996b, and in preparation), 
or an elliptical. Finally a lot can be learned from better 
modeling of nearby objects which still elude us, like M51 or the
Cartwheel.

The fate of globular clusters (GCs) during mergers can reveal a wealth of
information on the processes at work during the merging. Several
observations have now shown that the colour distribution 
of the GCs of many elliptical galaxies are bi-modal or even
multi-modal, arguing for the presence of more than one population of
GCs around the host galaxy. Several possibilities about their
formation have been discussed in the literature. Some
could have been initially attached to one of the spirals 
that merged to make the elliptical, while others could
have formed during the merger. Other GCs could have
initially formed in dwarf galaxies and been appropriated by the main 
elliptical during a minor merger. Fully
self-consistent high-resolution N-body simulations 
of mergings, both minor and major, in which the fate of the globular
clusters are followed with the help of realistic rules, are necessary
to understand the relative importance of the various origins proposed
above, as well as the spatial and velocity
distributions of the corresponding families of GCs.  
This study should be extended to galaxy clusters, where 
one has also to take into account that GCs can be tidally stripped from
their parent galaxies and accreted by the brightest 
cluster member. The wealth of recent observations on this subject are
well suited for comparisons with the results of N-body simulations.

More work is certainly necessary to understand the dynamical evolution 
of loose groups, and also under which (if any) conditions they can
lead to compact groups. This would shed more light on the question
whether observed compact groups are recently formed, or whether their
longevity is due to a massive and not centrally concentrated common halo
(Athanassoula, Makino \& Bosma 1997).

A deeper understanding of the dynamical evolution of galaxies which are part of
groups or clusters requires numerical simulations with a very high
number of particles. Except for a couple of notable exceptions,
so far progress has been achieved either by
simplifying the description of the galaxies (e.g. considering only
their halos), or by considering very small groups, or by assuming that the
cluster can be described by a rigid potential. All three have led to
some interesting results, although they have obvious shortcomings. Yet 
N-body simulations with a sufficient number of particles to describe a 
cluster of realistically modeled galaxies are, or will shortly be,
within the reach of several computers and progress should be fast in
this area.

Several observations of intra-group or intra-cluster stellar
populations exist (e.g. Freeman, these proceedings). Here again fully
self-consistent N-body simulations where each individual galaxy is
realistically modeled should shed some light on the origin and
evolution of debris. Some of my preliminary results on this subject
show that these should indirectly set constraints on the properties of 
the common halo of the group or cluster.

\subsection{Beyond pure stellar dynamics } \label{hydro+SF}

In order to model a particular phenomenon or effect it is sometimes
necessary to consider not only stars but also gas. 
The first question that arises in such cases
is how this gas should be modeled. Using hydrodynamic schemes based on
finite differences? Sticky particles? SPH? Before embarking into any
extensive use of gas in N-body simulations it seems necessary to compare
the results of the various methods of modeling gas in cases where
observations ``tell us the answer''. Thus in the case of the gas flow
in a rigid bar potential there is a very good agreement between SPH and
FS2 results (e.g. Patsis \& Athanassoula, in prep.) and a relatively good one
between FS2 and sticky particles
(Athanassoula \& Guivarch, in prep.), in as far as the response
morphology  is concerned. Similar work should be done to compare
the rate of gas inflow. The 
time-scale of the gas inflow depends on the properties
of the bar (mass, axial ratio, pattern speed, etc.), but also on
viscosity, and thus  
on the code, so that it is necessary to know how code dependent the
various estimates may be. 
It is thus important to compare inflow rates obtained with FS2, SPH,  
and other hydro approaches as well as include star
formation. Finally including multi-phase interstellar
medium might have still some surprises in store for us.

Star formation is on a yet more slippery ground. Various ``recipes''
have been used so far, based e.g. on Schmidt's law, or on Toomre's Q
parameter. One has also to take into account the feedback of the stars
on the gas, including heating by stellar winds and by supernovae. It
is clear that a tight collaboration with people working on star
formation would be most fruitful. Nevertheless the problem is rather
complicated and real progress may be expected to be slow, since
descriptions of numerous processes on a variety of spatial scales need 
to be combined in a unified framework.

\acknowledgments

I would like to thank A. Bosma and J. Makino for motivating
 discussions.


%
%

%

\end{document}